# Prediction of glass formability from liquid properties

R. Dai,[1,*] R. Ashcraft,[2] A. K. Gangopadhyay,[2] and K. F. Kelton[1,2]

[1]Institute of Materials Science and Engineering, Washington University in St. Louis, St. Louis, MO 63130, USA

[2]Department of Physics and Institute of Materials Science and Engineering, Washington University in St. Louis, St. Louis, MO 63130, USA

**Abstract**

Glass formation is one of the most interesting phenomena in the condensed matter field. Considerable effort has gone into understanding and predicting the glass formability. However, the previous prediction requires the glass first made before the prediction can be performed. Here, we propose a new prediction formula using liquid properties only. Moreover, we demonstrated that the similarity between liquid and crystalline structure plays an important role in determine the glass formability. Previously, only the kinetics of nucleation and growth processes have been considered.

## 1. Introduction

Metallic glasses have drawn significant attention due to their high strength, corrosion resistance, excellent elastic energy storage capacity and versatile processing capabilities[1]. They are formed when liquids are cooled fast enough to avoid crystallization. However, the key question as to why some liquids easily form metallic glasses while others do not is still unsolved.

Numerous works attempted to explain and predict glass formability (GFA) of metallic alloys, all having varying degrees of success[2–5]. Usually, the GFA is defined in terms of the critical casting thickness or the critical cooling rate for glass formation. Since glass formation is favored when crystal nucleation and growth from the liquid are avoided, the focus of these works was on factors that affect nucleation and growth kinetics. For example, Turnbull[6] suggested that metallic alloys with a reduced glass transition temperature, $T_{rg} = T_g/T_l$ ($T_l$ and $T_g$ are the liquidus and glass transition temperatures), greater than 2/3 are usually good glass formers. The reasoning is that with larger $T_{rg}$, ($T_l - T_g$)/$T_l$ will be smaller. A smaller temperature window between $T_l$ and $T_g$ decreases the driving free energy for crystal nucleation and, therefore, decreases the probability of crystal nucleation. However, $T_{rg}$ alone is an inadequate quantitative predictor of glass formability[2]. It has been argued that a combination of the kinetic fragility parameter, $m = \frac{d(log_{10}(\eta))}{d(T_g/T)}$ at $T_g$, and $T_{rg}$ may be better indicators for GFA [2,4]. Here, the reasoning is that a more viscous equilibrium and metastable liquid (below $T_l$) slows down atomic mobility and, therefore, decreases both nucleation and growth rates. A combination of slower atomic mobility and smaller temperature window for the liquid state (larger $T_{rg}$) then favor glass formation.

However, $m$ is not easy to measure and the reported $m$ values often have substantial disparities[7]. For example, the reported $m$ values[8–10] range from 52 to 109 for $Pd_{77.5}Cu_6Si_{16.5}$. Furthermore, $m$ is difficult to measure for marginal glass formers because of rapid crystallization as $T_g$ is approached. Most importantly, to predict GFA according to the above scheme, the glass has to be made first, the above properties have to be measured, and then only one can predict glass formability, which is a self-defeating process. The natural question is, can GFA be predicted without making a glass? Is it possible to predict GFA from the properties of the equilibrium liquid? Is there a new search algorithm for good glass formers that does not require knowing $T_g$ in advance? As a part of these broader questions, we recently introduced a method to predict $T_g$ from the high-temperature viscosity and liquid expansion coefficient (α) [11]. It has also been shown [12,13] that it is possible to define fragility in terms of high temperature viscosity. If a temperature $T^*$ is defined for a specific value of viscosity for all liquids, then $T_g/T^*$ is a good measure of liquid fragility. Therefore, instead of using measured $m$ and $T_{rg}$ [2,4], it is possible to define GFA in terms of experimentally determined $T^*$, $T_l$ and predicted $T_g$ from α and high-temperature viscosity [11]. All of these parameters are measured for the equilibrium liquid. A search algorithm is then used to predict the critical thickness in terms of the above parameters and compared with the *literature data*. As will be shown, the agreements are reasonable, considering that the literature data for the critical thickness usually show large variations with microscopic impurities (often oxides) in the sample.

However, in a few cases complete failure of the prediction was observed. For example, $Zr_{80}Pt_{20}$ was predicted to form a glass with a critical thickness of 10 mm. In practice, it is one of the poorest glass formers[14]. This indicates that some additional factors are in play. In the classical nucleation theory[15], other than the driving free energy and viscosity/diffusivity, an important parameter that determines the nucleation rate is the interfacial energy between the solid nucleus and the surrounding liquid. It was demonstrated quite a while ago[16] that the local orders in the liquid and the crystal phase are very important factors that determine interfacial free energy; if the short/medium range order in the crystal and liquid phases are similar, the interfacial energy becomes small and the crystal nucleation becomes much easier. A systematic study of the structures of the liquid and crystal phases for the anomalous cases in the present study shows that the failure of the prediction comes from the absence of this information in the predictive theory. This is the first demonstration of the importance of this missing parameter that needs to be incorporated in any successful theory for the GFA; thus far, only the kinetics of nucleation and growth processes have been considered.

## 2. Experimental Methods

Most of the data used in the present analysis were reported earlier in some form or other [11,12]. Therefore, details of the experimental methods can be found in the earlier reports. Briefly, small samples (30-60 mg) for the electrostatic levitation (ESL [17,18]) studies were prepared from larger ingots (~1g) by arc-melting of high purity elements. The viscosity and thermal expansion coefficients of the equilibrium and supercooled (below $T_l$) liquids were measured on molten levitated droplets under high vacuum (~$10^{-8}$ Torr). Density and thermal expansion coefficients were obtained from the temperature dependence of volume, determined from the image-analysis [19–21] of the spherical droplets. The viscosities were determined from the decay of induced surface oscillations of the droplets using standard procedures [22–26].

## 3. Results and discussions

Using $T_g/T^*$ as the fragility index, a prediction formula for critical casting thickness ($d_{max}$) can be developed in terms of $T_{rg}$ and $T_g/T^*$ following Johnson et al. [2]. Table 1 contains a summary of experimental $T_{rg}$, $T_g/T^*$, and $d_{max}$ for 15 existing glass formers [11,12]. The largest $d_{max}$ reported in the literature are used since they are probably the most representative of the intrinsic glass formability of the alloys; smaller $d_{max}$ are probably due to sample contamination. However, they were reported from different groups without an assessment in a systematic manner and the error of $d_{max}$ was estimated to be around 15%[2]. The least-square fit of all data provides an empirical relationship:

$$\log(d_{max}^2) = 7.232 + 13.629\, T_{rg} - 22.896 T_g/T^* , \qquad (1)$$

Equation 1 suggests that alloys with higher $T_{rg}$ and smaller $T_g/T^*$ (stronger liquids) are better glass formers, which agrees with other studies[2].

As mentioned in Ref [11], $T_g$ can be accurately predicted using high temperature viscosity and liquid expansion coefficient data, as shown in equation 2.

$$T_g = \frac{(1920*\alpha + 0.297)}{(1/T^* - 0.307/T_A)}. \tag{2}$$

where $\alpha$ is the thermal expansion coefficient, $T^*$ is the temperature at which the viscosity reaches a common value(0.06 Pa-s) for all studied liquids, $T_A$ is the temperature below which the shear viscosity becomes super-Arrhenius[7,27].

Table 1 also includes experimental and predicted $T_g$ values. The difference between them is between 0.2% and 3.5%, with an average deviation of 1.5%.

By replacing experimental $T_g$ with predicted $T_g$, a truly predictive glass formability model from liquid data can then be proposed.

$$\log(d_{max}^2) = 7.232 + \frac{26167.68*\alpha + 4.048}{T_l/T^* - 0.307*T_l/T_A} - \frac{43960.32*\alpha + 6.8}{1 - 0.307*T^*/T_A}, \tag{3}$$

The experimental $d_{max}$ versus predicted $d_{max}$ using equation 3 are shown in Table 1.

**Table 1**. The experimental $d_{max}$, $T_g$ [11], $T_{rg}$ [11,12], $T_g/T^*$ [11,12], predicted $T_g$ [11], and predicted *dmax* values for 15 existing glass formers.

| Composition | dmax (mm) | Experimental $T_{rg}$ | Experimental $T_g/T^*$ | Experimental $T_g$ (K) | Predicted $T_g$ (K) | Predicted dmax (mm) |
|---|---|---|---|---|---|---|
| Cu46Zr54 | 2[28] | 0.5484 | 0.6083 | 657 | 638.1 | 3 |
| Cu47Zr47Al6 | 6[29] | 0.5913 | 0.6068 | 693 | 691.3 | 5 |
| Cu50Zr40Ti10 | 4[30] | 0.5616 | 0.5899 | 656 | 660.4 | 5 |
| Cu50Zr42.5Ti7.5 | 5[31] | 0.5807 | 0.6270 | 669 | 645.5 | 3 |
| Cu50Zr45Al5 | 3[32] | 0.5925 | 0.6134 | 695 | 688.7 | 5 |
| Cu50Zr50 | 2[33] | 0.5450 | 0.6011 | 666 | 659.1 | 3 |
| Cu64Zr36 | 2[34] | 0.6167 | 0.6584 | 740 | 723.4 | 2 |
| Ti40Zr10Cu30Pd20 | 3[35] | 0.5635 | 0.5993 | 670 | 687.2 | 3 |
| Vit105 | 18[2] | 0.6139 | 0.5536 | 671 | 681.1 | 27 |
| Vit106 | 20[2] | 0.5975 | 0.5514 | 671 | 669.3 | 24 |
| Vit106a | 32[2] | 0.5938 | 0.5590 | 668 | 664.6 | 19 |
| Zr56Co28Al16 | 18[36] | 0.5955 | 0.5611 | 739 | 742 | 17 |
| Zr60Ni25Al15 | 15[37] | 0.5565 | 0.5556 | 694.5 | 682 | 12 |
| Zr64Ni25Al11 | 12[37] | 0.5520 | 0.5475 | 669 | 686.3 | 11 |
| Zr65Al7.5Cu17.5Ni10 | 16[38] | 0.5470 | 0.5479 | 640 | 650 | 11 |

As seen from Table 1, the predicted and literature data for $d_{max}$ are in fair agreement except for Vit alloys. The predicted critical casting thickness for Vit 106 (24 mm) is larger than that of Vit 106a (19 mm), which is in contradiction with the experimental results[39,40]. Vit106 has a larger $T_{rg}$ and smaller $T_g/T^*$ (kinetically stronger) than Vit106a. This should suggest a larger $d_{max}$ for Vit106 compared to Vit106a, as predicted. As we will discuss later, the failure of the prediction is because we didn't include one important predictor. Some other compositions not in the training data set were used to verify the prediction (See Table 2). First, we focus on $Zr_{62}Cu_{20}Ni_8Al_{10}$ and $Zr_{59}Ti_3Ni_8Cu_{20}Al_{10}$. Compared to the predicted values of 14 mm and 13 mm for the two alloys, the reported $d_{max}$ are only 3 mm. Interestingly, based on containerless solidification experiments, Kim et al. [41] reported that $Zr_{62}Cu_{20}Ni_8Al_{10}$ has a lower critical cooling rate for glass formation, compared with a good glass former $Zr_{57}Ti_5Ni_8Cu_{20}Al_{10}$. Therefore, the critical casting thickness for $Zr_{62}Cu_{20}Ni_8Al_{10}$ should be larger than that of $Zr_{57}Ti_5Ni_8Cu_{20}Al_{10}$. Since 10 mm diameter amorphous samples can be produced by melt-casting of $Zr_{57}Ti_5Ni_8Cu_{20}Al_{10}$ [42], the $d_{max}$ for $Zr_{62}Cu_{20}Ni_8Al_{10}$ should be larger than 10 mm, as predicted. Using the same containerless levitation technique, we have observed that $Zr_{59}Ti_3Ni_8Cu_{20}Al_{10}$ can also form a glass during free-radiation cooling in the ESL, while $Zr_{57}Ti_5Ni_8Cu_{20}Al_{10}$ does not. This suggests that the $d_{max}$ for $Zr_{59}Ti_3Ni_8Cu_{20}Al_{10}$ should also be larger than 10 mm, again consistent with prediction. Therefore, the anomalously small experimental $d_{max}$ for these two alloys is most likely due to microscopic contamination of the samples during processing.

Since there is no reported $T_g$ for $Zr_{80}Pt_{20}$, its $T_g$ was either estimated from the onset of crystallization[12] or predicted from the liquid data[11]. The predicted $d_{max}$ for $Zr_{80}Pt_{20}$ are 17 mm and 10 mm respectively. In contrast, glass formation in this alloy has never been reported; even in melt-spinning experiments using Cu-wheel, icosahedral quasicrystal formation, instead of a glass, was reported[43]. Similar observations were made by us even when this alloy was quenched with a 70 m/s wheel speed. Therefore, the critical casting thickness for this alloy must be less than 10 μm, compared to the predicted value. The predicted $d_{max}$ for $Cu_{43}Al_{12}Zr_{45}$ is 9 mm compared to 7 mm for $Cu_{47}Al_8Zr_{45}$, according to eqn. (3). Experimentally, the reported $d_{max}$ for the $Cu_{47}Al_8Zr_{45}$ glass is 15 mm[44], whereas no data for the critical casting thickness of $Cu_{43}Al_{12}Zr_{45}$ is available. To further check the prediction, copper mold casting technology was used to synthesize $Cu_{43}Al_{12}Zr_{45}$ and $Cu_{47}Al_8Zr_{45}$ glasses. However, compared to a fully amorphous structure for a 1.6 mm thick $Cu_{47}Al_8Zr_{45}$, the $Cu_{43}Al_{12}Zr_{45}$ sample of the same thickness is crystalline. This demonstrates that the GFA of $Cu_{47}Al_8Zr_{45}$ must be better than $Cu_{43}Al_{12}Zr_{45}$, which is in conflict with predictions.

To understand these discrepancies, we now draw the attention to the SRO in the liquids and corresponding crystals. Mauro et al. [45] reported a significant amount of icosahedral order in $Zr_{80}Pt_{20}$ liquid; the corresponding crystal phase is also an icosahedral quasicrystal as mentioned earlier[43]. Evenson et al.[46] have suggested that compared with Vit106a, the local order in Vit106 liquid is more similar to that of the primary crystalline phase. These results then suggest that the failure of the prediction occurs when the liquid and crystalline phases have very similar structural order. This observation is quite consistent with a conclusion made a decade ago from studies of Ti-Zr-Ni liquids[16]. It was demonstrated that when the SRO in the liquids and crystals are very similar, nucleation of the crystal phase becomes much easier. The underlying order in the liquid acts as a template for easy nucleation (almost like a heterogeneous nucleation) of the crystal. Such an easy nucleation pathway completely overwhelms the tendencies to avoid crystallization even when the liquid is strong and $T_{rg}$ is large. To further provide support for the above observation, we examined the liquid and crystal structures of $Cu_{47}Al_8Zr_{45}$ and $Cu_{43}Al_{12}Zr_{45}$ for the first time in great detail.

**Table 2**. The experimental $d_{max}$, $T_g$ [11], $T_{rg}$ [11,12], $T_g/T^*$ [11,12], predicted $T_g$ [11], and predicted $dmax$ values for compositions not in the training data set. The blank means no literature data were reported.

| Composition | $dmax$ (mm) | Experimental $T_{rg}$ | Experimental $T_g/T^*$ | Experimental $T_g$ (K) | Predicted $T_g$ (K) | Predicted $dmax$ (mm) |
|---|---|---|---|---|---|---|
| Zr59Ti3Ni8Cu20Al10 | 3[47] | 0.5712 | 0.5657 | 654 | 631 | 13 |
| Zr62Cu20Ni8Al10 | 3[47] | 0.5686 | 0.5560 | 655 | 650 | 14 |
| Zr80Pt20 | | 0.4903[a] | 0.5011[a] | 710[a] | | 17[a] |
| Zr80Pt20 | | | | | 772.4 | 10 |
| Cu43Zr45Al12 | | 0.6236 | 0.6038 | 724 | 722.2 | 9 |
| Cu47Zr45Al8 | 15[44] | 0.6071 | 0.6045 | 706 | 705.3 | 7 |

[a] $T_g$ was estimated from the crystallization onset temperature.

Crystalline structures of $Cu_{43}Al_{12}Zr_{45}$ and $Cu_{47}Al_8Zr_{45}$ were investigated using the ESL-based x-ray diffraction facility (BESL). The temperature-time profile of the $Cu_{43}Al_{12}Zr_{45}$ liquid is plotted in Figure 1 as the sample was cooled from much above $T_l$ (340K) by radiative heat loss. Several recalescence events (sudden rise in temperature) were observed. During the first recalescence, the liquid partially crystallized. Then the remaining liquid continues to crystallize until around 301 s, when the crystallization completed. *In situ* x-ray diffraction patterns (Figure 2, top) shows that after the first recalescence, the remaining liquid crystallized into a different structure during the 2$^{nd}$ recalescence. GSAS II [48] was used to identify the crystalline structures. The crystal formed during the first recalescence is identified to be AlCuZr (Laves(cub)-Cu$_2$Mg like); the corresponding peaks are labeled by red stars in Figure 2. Fully crystallized structure was identified to be a mixture of CuZr, AlCuZr (Laves(cub)-Cu$_2$Mg like), and Al$_2$Zr (Figure 2, bottom). A Voronoi analysis [48] of the AlCuZr phase shows that the local symmetry is dominated by perfect icosahedral clusters; <0 0 12 0> index belongs to 53% and <0 0 12 4> polyhedra belongs to 26%. A common neighbor analysis(CNA) conducted by Bailey et al.[49] also found that the Cu atoms have CN12 icosahedral coordination while the Mg atoms are surrounded by CN 16 pylyhedra in Cu$_2$Mg . It is therefore not surprising that <0 0 12 0> and <0 0 12 4> indices are prevalent in the Laves(cub)-Cu$_2$Mg like AlCuZr phase.

In contrast, only one recalesence event marks the crystallization from supercooled Cu$_{47}$Al$_8$Zr$_{45}$ liquid (Figure 3, top). The crystalline structure was identified to be a mixture of BCC CuZr and Al$_1$Cu$_2$Zr$_1$ (see Figure 3, bottom). From the Voronoi analysis, both of these crystal types are dominated by <0 6 0 8> clusters.

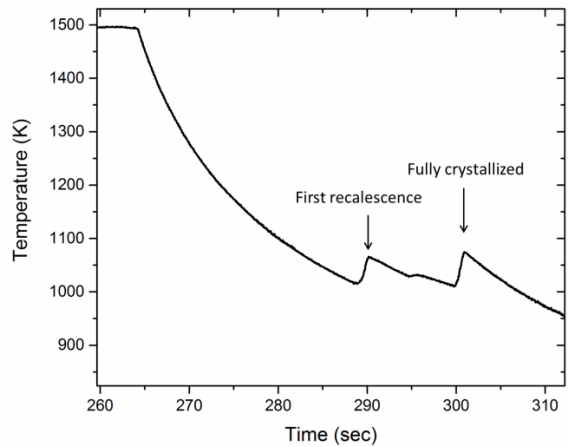

Figure 1. Temperature-time profile for a Cu$_{43}$Al$_{12}$Zr$_{45}$ levitated droplet during free radiation cooling in the BESL.

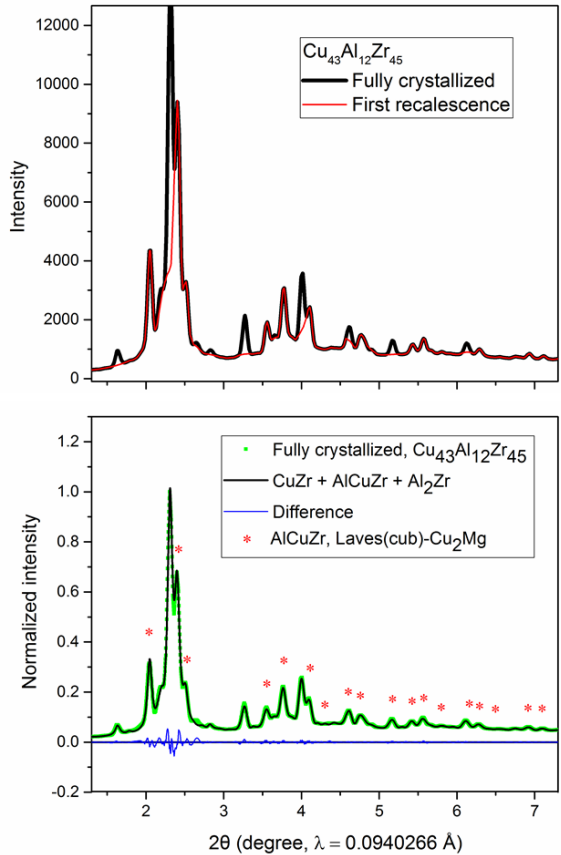

Figure 2. (Top) *In situ* x-ray diffraction patterns for $Cu_{43}Al_{12}Zr_{45}$ liquid after full crystallization (black) and first recalescence (red). (Bottom) Crystalline phase identification of measured intensity after fully crystallization (green points) and model fits (black line) scaled to the maximum intensities. The difference between the measured intensity and the fit is shown below the fit (blue line). The peaks corresponding to the structure after the first recalescence belongs to AlCuZr (Laves(cub)-$Cu_2Mg$ like, red stars).

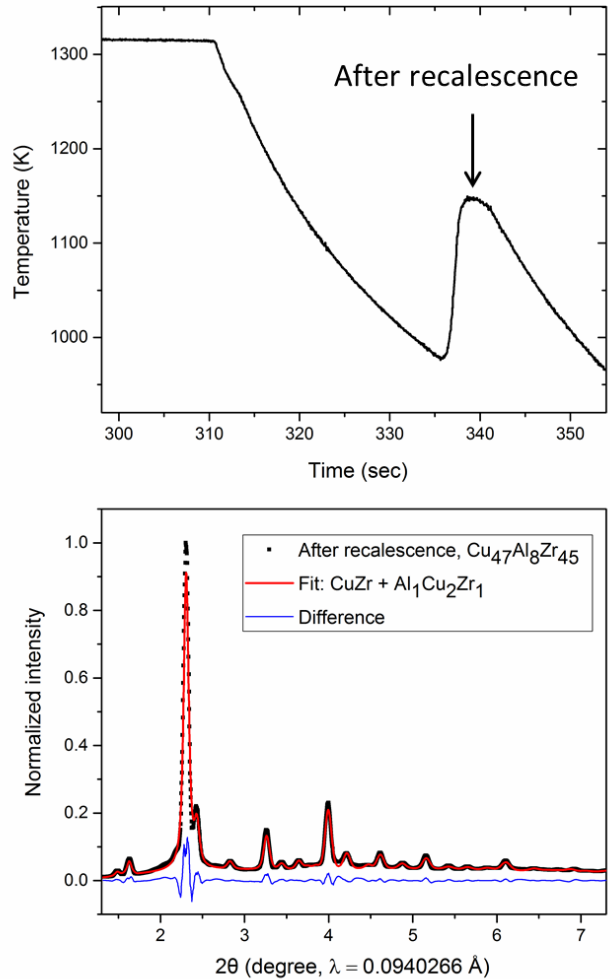

Figure 3. (Top) Free cooling curve for $Cu_{47}Al_8Zr_{45}$ liquid showing one recalescence. (Bottom) Structure identification for $Cu_{47}Al_8Zr_{45}$ with measured intensity after recalescence (black points) and model fits (red line) scaled to the maximum intensities. A solid bule line shows the difference.

Similarly, the structures of $Cu_{43}Al_{12}Zr_{45}$ and $Cu_{47}Al_8Zr_{45}$ liquids were also examined. Reverse Monte Carlo (RMC) fits, constrained with the X-ray static structure factors, $S(q)$, were used to gain information about the average topology of the atomic configurations. Each simulation was run for 60 hours until the difference between the RMC fits and the experimental data was minimized. 10 simulations were conducted for each temperature of interest to the extract the statistical information of the liquid structure. RMC configurations were then analyzed using Voronoi tessellation techniques[50,51]. Particular attention was focused on <0 0 12 0> and <0 0 12 4> clusters for $Cu_{43}Al_{12}Zr_{45}$ liquid and <0 6 0 8> clusters for $Cu_{47}Al_8Zr_{45}$ liquid because those clusters are the dominant ones in the corresponding crystalline structures. The number of <0 0 12 0> and <0 0 12 4> clusters increase with decreasing temperatures, as shown in Figure 4 (a) and (c) for the $Cu_{43}Al_{12}Zr_{45}$ liquid; however, the relative number for the 2$^{nd}$ type is much smaller. The $Cu_{43}Al_{12}Zr_{45}$ liquid also contains many distorted icosahedra, such as <0 2 8 2>, which also increases with decreasing temperatures, as shown Figure 4 (b). These results indicate that the atomic configurations in $Cu_{43}Al_{12}Zr_{45}$ liquid evolve more towards the SRO of the underlying crystal phase with decreasing temperatures. This facilitates easy nucleation of the crystal, which is detrimental for glass formation. In contrast, the <0 6 0 8> cluster type doesn't show any significant increase with decreasing temperature in the $Cu_{47}Al_8Zr_{45}$ liquid. Therefore, the SRO in the liquid does not show a trend towards developing an order similar to that in the underlying crystal phase. As a result, glass formation becomes easier for this composition. This provides a natural explanation why the actual $d_{max}$ is larger in $Cu_{47}Al_8Zr_{45}$ although from the fragility and $T_{rg}$ considerations $Cu_{43}Al_{12}Zr_{45}$ should be a better glass former.

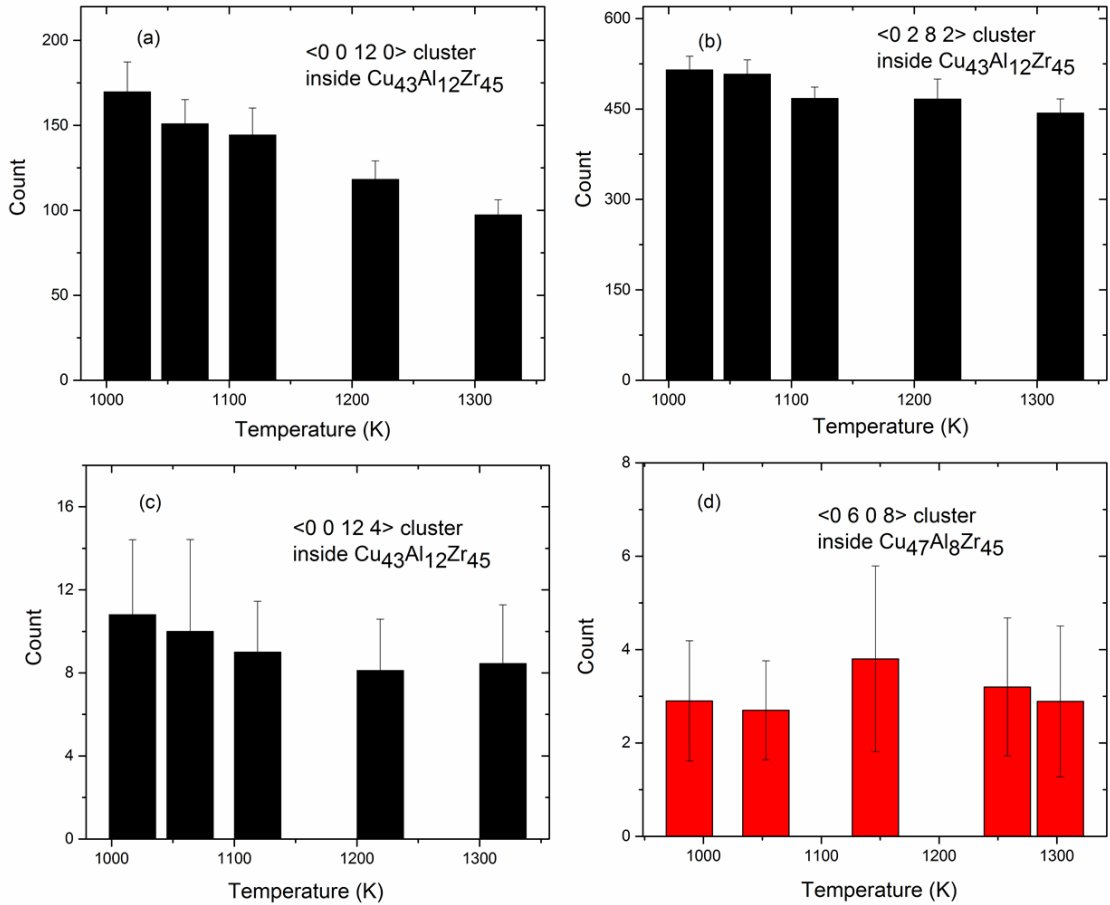

Figure 4. The number of (a) <0 0 12 0>, (b) <0 2 8 2>, (c) <0 0 12 4> clusters for $Cu_{43}Al_{12}Zr_{45}$ liquid as temperature decreases. (d) <0 6 0 8> cluster is negligible in $Cu_{47}Al_8Zr_{45}$ liquid and doesn't increase its number with decreasing temperature.

## Summary and Conclusions

In summary, a new method of predicting GFA of metallic glasses from the properties of the equilibrium liquids has been presented. The necessary input parameters are the liquidus, viscosity, and thermal expansion coefficients of the equilibrium liquids. Although the predicted $d_{max}$ are in agreement with the literature data in most cases, some anomalies were also noticed. It is argued in some cases, such as $Zr_{62}Cu_{20}Ni_8Al_{10}$ and $Zr_{59}Ti_3Ni_8Cu_{20}Al_{10}$, that the predictions are correct and the experimental $d_{max}$ are possibly underestimated due to microscopic amounts of impurities in the samples. However, this could not be verified experimentally because of lack of experimental facilities available to this group. In other cases, such as $Zr_{80}Pt_{20}$, Vit106, and $Cu_{43}Al_{12}Zr_{45}$ and $Cu_{47}Al_8Zr_{45}$ alloys, a new paradigm has been identified that contributes to glass formation over and above the parameters such as fragility and reduced glass transition temperatures used in the present and some previous models[2,4]. This new paradigm is that when the SRO in the crystal and liquids are similar, glass formation becomes difficult; when they are dissimilar, GFA is enhanced. Although this criterion can be evaluated from analyzing the liquid and nucleating crystal structures from *in-situ* diffraction experiments without synthesizing the glass, it is not clear how a quantitative parameter may be introduced in the model to include this important effect.

## Acknowledgments

The authors thank Doug Robinson for his assistance with the high-energy X-ray diffraction measurements at the Advanced Photon Source (APS), Ryan J Chang for making and testing $Cu_{47}Al_8Zr_{45}$ and $Cu_{43}Al_{12}Zr_{45}$ samples, and Vy Tran and Ryan Soklaski for the use of their Voronoi code. This work was partially supported by the National Science Foundation under grant **** and NASA under grant ***. The synchrotron measurements were made on the Sector 6 beamline at the APS. Use of the APS is supported by the US Department of Energy, Basic Energy Science, Office of Science, under contract no.****. Any opinions, findings, and conclusions or recommendations expressed in this publication are those of the authors and do not necessarily reflect the views of the National Science Foundation or of NASA.